# A Source-Independent Fault Detection Method for Transmission Lines

Reza Jalilzadeh Hamidi, *Senior Member, IEEE* and Julio Rodriguez, *Student Member, IEEE*

*Abstract*—This article proposes a source-independent method for detecting faults along Transmission Lines (TL) to reduce the protection issues arising from Inverter-Based Resources (IBRs). In the proposed method, high-frequency waves are sent from either end of a TL, and the amplitudes of the receiving waves at the other end are measured. Faults change the characteristics of TLs. Therefore, the amplitudes of the receiving waves differ when a fault occurs. Closed-form formulations are developed for describing the receiving waves before and during the faults. These formulations indicate that at least one of the receiving waves is reduced after fault inception. Therefore, faults can be detected by identifying a decrease in one of the receiving waves.

To evaluate the performance of the proposed method, EMTP-RV is utilized for performing simulations. Additionally, laboratory experiments are conducted for further evaluation of the proposed method. The simulation and experimental results demonstrate that the proposed method is able to detect faults along TLs regardless of the sources supplying the grid.

## I. INTRODUCTION

THE integration of Inverter-Based Resources (IBRs) into power grids has experienced a significant surge. Despite the benefits of IBRs, they disturb the performance of conventional protective devices (i.e., phasor-based relays). The relays have been developed premised on the fault responses of Electro-Mechanical Generators (EMGs). However, the fault responses of IBRs are drastically different compared to those of EMGs [1] and [2]. Several methods have been proposed in the literature to address the protection issues arising from IBRs, as follows:

One solution is to emulate the fault responses of EMGs by IBRs [3]. However, it is challenging for IBRs to generate high fault currents mainly due to their thermal limits [1]. It is more challenging for IBRs with uncontrollable primary energy sources (e.g., solar panels). Since it is probable that the primary energy source is insufficient for generating high enough fault currents [1]. For removing this barrier, the addition of Energy Storage Systems (ESSs), such as supercapacitors, is suggested in [4]. Despite the effectiveness of this solution, integrating ESSs into IBRs is economically challenging.

The other solution is that IBRs increase their positive- and negative-sequence reactive currents during faults [5]-[7]. This method is successful in mitigating the adverse effects of IBRs on relays. However, this solution is aimed at the cooperation of EMGs and IBRs, and it is less effective in fully IBR-supplied grids [1] and [6].

The adoption of fault-induced Travelling Waves (TWs) for fault detection is another remedy [8] and [9]. However, the performance of TW-based solutions is dependent on the grid characteristics, pre-fault conditions, fault characteristics, Fault Inception Angle (FIA), and the sampling rate of the measuring devices [10] and [11]. Therefore, the application of TW-based methods is practically challenging.

Following adaptive protection schemes, the settings of relays are continually adjusted according to the grid state [12] and [13]. Adaptive protection schemes heavily rely on communications. Therefore, they are susceptible to cyber-attacks and communication failures [4] and [14].

The development of relays based on incremental quantities is another solution [15] and [16]. Although the use of incremental quantities is effective in detection of faults in IBR-based grids, the quantities are heavily dependent on the grid parameters. Therefore, precise models and parameters of grids are required that reduces the applications of incremental quantities-based relays [17].

The application of Artificial Intelligence (AI) is another solution to overcoming the IBR-caused protection challenges [18] and [19]. Although high accuracy in fault detection using AI-based methods is reported, their non-deterministic results and need for extensive datasets curtail their suitability as a primary protection scheme [20].

In [21], High-Frequency (HF) waves generated by Power Line Carriers (PLCs) are employed for detecting high-impedance faults in distribution grids. This method determines the network impedance against the HF waves. Then, unexpected changes in the determined impedance can indicate the occurrence of faults. However, this method requires that all the TL ends are open. In addition, transformers can filter HF waves. Thus, the installation of equipment for bypassing transformers is necessary.

According to the abovementioned shortcomings in detecting faults in IBR-energized grids, this paper proposes a source-independent fault detection method for TLs. The importance of the proposed method is rooted in that it increases the hosting capacity of grids for IBRs.

Reza Jalilzadeh Hamidi and Julio Rodriguez are with the Department of Electrical and Computer Engineering, Georgia Southern University, Statesboro, GA 30458, USA (e-mails: reza.j.hamidi@gmail.com; jr33130@georgiasouthern.edu).

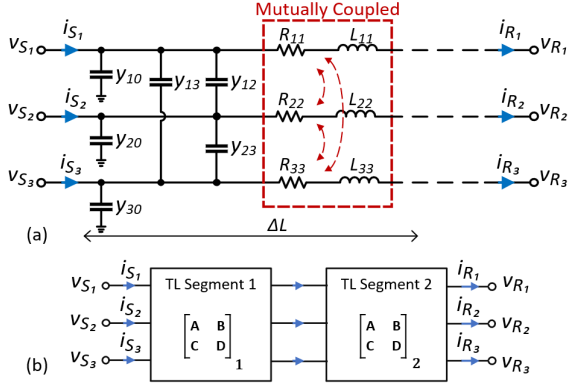

**Fig. 1.** (a) The distributed circuit model of three-phase TLs. (b) The presentation of a TL with two sequential segments.

The proposed method adopts the principles of wave propagation in TLs, in that HF waves are sent from either end of a TL. At the same time, the receiving waves are measured at the other ends. As faults change the features of the TL, the amplitudes of the receiving waves are changed by faults. Accordingly, the proposed method recognizes faults by identifying the changes in the amplitudes of the receiving waves. The key features of the proposed method are: 1) The proposed method does not require the line parameters. 2) It is able to detect faults with relatively high impedances. 3) The proposed method is able to detect faults in de-energized TLs. This helps maintenance crews ensure that the faults are cleared before re-energizing the lines. The simulation and experimental results demonstrate that the proposed method is able to effectively detect the faults along TLs.

II. PRINCIPLES OF THE PROPOSED METHOD

In this section, first, the model of TLs is reviewed, and then the requirements for the proposed method are discussed.

*A. Three-Phase TL Model*

The distributed circuit model of three-phase TLs is illustrated in Fig. 1(a). The relations between the sending and receiving voltages and currents of TLs are [22]

$$\begin{bmatrix} V_S \\ I_S \end{bmatrix}_{6 \times 1} = \pi_{6 \times 6} \begin{bmatrix} V_R \\ I_R \end{bmatrix}_{6 \times 1} \quad (1)$$

where $V_S = [v_{S_1}, v_{S_2}, v_{S_3}]^T$ and $V_R = [v_{R_1}, v_{R_2}, v_{R_3}]^T$ are respectively the vectors of the sending and receiving phase-to-ground phasor voltages. $I_S = [i_{S_1}, i_{S_2}, i_{S_3}]^T$ and $I_R = [i_{R_1}, i_{R_2}, i_{R_3}]^T$ denote the vectors of sending and receiving phasor currents, respectively. The elements of the so-called pi-matrix representing TLs are [22]

$$\pi = \begin{bmatrix} A & B \\ C & D \end{bmatrix}$$
$$= \begin{bmatrix} [\cosh(\gamma l)]_{3 \times 3} & [\sinh(\gamma l) Z_c]_{3 \times 3} \\ [Z_c^{-1} \sinh(\gamma l)]_{3 \times 3} & [Z_c^{-1} \cosh(\gamma l) Z_c]_{3 \times 3} \end{bmatrix} \quad (2)$$

where $\gamma^2 = zy$, $Z_c = \gamma^{-1} z$, and $z$ and $y$ are per length unit impedance and admittance matrices of the TL, as follows:

$$z = \begin{bmatrix} z_{11} & z_{12} & z_{13} \\ z_{21} & z_{22} & z_{23} \\ z_{31} & z_{32} & z_{33} \end{bmatrix}, y = \begin{bmatrix} y_{11} & y_{12} & y_{13} \\ y_{21} & y_{22} & y_{23} \\ y_{31} & y_{32} & y_{33} \end{bmatrix} \quad (3)$$

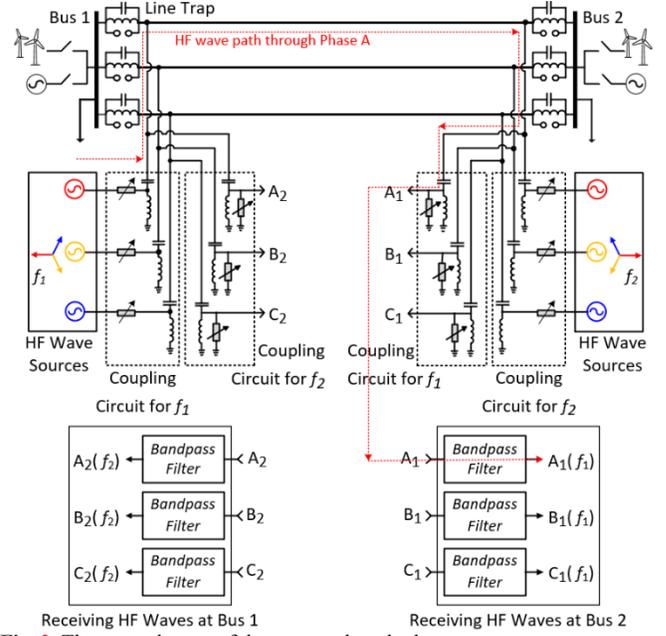

**Fig. 2.** The general setup of the proposed method.

where $z_{ij}$ is the mutual impedance per length unit between Phases $i$ and $j$ if $i \neq j$, or the self-impedance per length of Phase $i$ if $i = j$. $y_{ij}$ is the negative of the admittance per length unit between Phases $i$ and $j$ if $i \neq j$, but when $i = j$, $y_{11} = y_{10} + y_{12} + y_{13}$, $y_{22} = y_{12} + y_{20} + y_{23}$, and $y_{33} = y_{13} + y_{23} + y_{30}$ in which $y_{10}$, $y_{20}$, and $y_{30}$ are phase-to-ground admittances per length unit. The parameters of the pi-matrix ($\pi$) can calculated, as follows [23] and [24]:

$$\begin{cases} A = I + \frac{(zy)l^2}{2!} + \frac{(zy)^2 l^4}{4!} + \cdots \\ B = \left( I + \frac{(zy)l^2}{3!} + \frac{(zy)^2 l^4}{5!} + \cdots \right) zl \\ C = yl \left( I + \frac{(zy)l^2}{3!} + \frac{(zy)^2 l^4}{5!} + \cdots \right) \\ D = z^{-1} \left( I + \frac{(zy)l^2}{2!} + \frac{(zy)^2 l^4}{4!} + \cdots \right) z \end{cases} \quad (4)$$

where $I$ is the identity matrix, and $l$ is the length of the TL. A TL can be divided into two sequential segments, as shown in Fig. 1(b). The parameters of the matrices representing the segments must be adjusted according to their lengths.

*B. Requirements of the Proposed Method*

Referring to Fig. 2, in the proposed method, three HF sinusoidal waves with the same amplitudes, but with 120° phase displacement, are continuously sent from either end of a TL. Therefore, two sets of balanced phasors, but at discernably different frequencies of $f_1$ and $f_2$, are traveling along the TL in opposing directions. To this end, components similar to those necessary for the interconnection of PLCs are required.

As shown in Fig. 2, two coupling circuits, together with impedance matching units, are required at each end of the line [25]. As the frequencies of the sending and receiving waves are different, one coupling circuit is for sending the waves, and the other one is for receiving the waves. The impedance matching units prevent the formation of standing waves. Line traps are also installed at both ends of the TL for restricting the HF waves within the TL [25]. Bandpass filters are required for removing the noises and wave interferences from the receiving waves.

The sent HF waves propagate along the TL and reach the other end of the TL. The receiving waves at Bus 2 on Phases A, B, and C, which are coming from Bus 1, are respectively indicated with $A_1$, $B_1$, and $C_1$ in Fig. 2. Similarly, the receiving waves at Bus 1 on Phases A, B, and C, which are coming from Bus 2, are denoted with $A_2$, $B_2$, and $C_2$ in Fig. 2.

## III. THE PROPAGATION OF HF WAVES DURING FAULTS

### A. Qualitative Description

Fig. 3 depicts the lattice diagram of HF waves propagating along the phases of a TL during an SLG fault. $v_{fa11}$, $v_{fb11}$, and $v_{fc11}$ are the forward voltage waves at the sending end (i.e., Location 1 in Fig. 3). The forward waves travel along the line and reach the fault location (i.e., Location 2 in Fig. 3). Although only Phase A is faulty, all three forward waves break at the fault location since phases are mutually coupled. Therefore, the discontinuity on Phase A affects the other waves to a lesser extent. The forward wave on Phase A ($v_{fa12}$) breaks into three parts at the fault location: one part passes through the fault impedance, another part reflects to the sending end ($v_{ra12}$), and the other part ($v_{fa22}$) continuous traveling and reaches the receiving end of the TL ($v_{fa23}$). Since tuning impedances are installed at both sides of the TL, the waves do not reflect at the sending (Location 1) or receiving (Location 3) ends.

Fig. 4 depicts typical propagations of 50-kHz waves in the case of an SLG fault at 20 km from the sending end of a balanced 100-km TL.

Three forward waves (Forward Wave 1 in Fig. 4) with the same amplitude of 1000 V, but with 120° phase displacement, start traveling from the sending end. When the waves reach the fault location, they break. The reverse waves (Reverse Wave 1 in Fig. 4) return to the sending end. The combination of Forward Wave 1 and Reverse Wave 1 generates a standing wave in the first segment of the TL. Two antinodes are generated per wavelength. The wavelength of the traveling waves is almost 6 km. Therefore, it is expected that almost seven antinodes are formed in 20 km. It can be observed in Fig. 4 that the number of antinodes is seven.

In addition, the voltage at the fault location (i.e., fault voltage) is the summation of the Forward Wave 1 and Reverse Wave 1. The voltage of the forward wave on Phase A is $958.06\angle -128.13$ V, and the voltage of Reverse Wave 1 is $771.60\angle 51.65$ V. Therefore, the fault voltage is their vectorial sum, which is $186.48\angle -127.27$ V.

It is clear that the forward waves are exponentially attenuated in Segment 1. However, the forward waves in Segment 2 (i.e., Forward Wave 2) are not exponentially attenuated since they are no longer balanced and have additive or subtractive effects on one another.

### B. Mathematical Description

With reference to Fig. 1(b), a TL can be divided into two segments. Segment 1 is between the sending end and fault location. Segment 2 is the rest of the line. The voltages and currents at the ends of Segment 1 are linked by

$$\begin{bmatrix} V_{f11} \\ I_{f11} \end{bmatrix} = \pi_1 \begin{bmatrix} V_{f12} \\ I_{f12} \end{bmatrix} \quad (5)$$

where, with reference to Fig. 3, $V_{f11} = [v_{fa11}, v_{fb11}, v_{fc11}]^T$ and $I_{f11}$ are the vectors of the forward voltage and current waves at the sending end of Segment 1, the elements of $\pi_1$ are adjusted according to the length of Segment 1, $V_{f12} = [v_{fa12}, v_{fb12}, v_{fc12}]^T$ and $I_{f12}$ are the vectors of forward voltage and current waves at the end of Segment 1. Since $I_{f11} = Y_c V_{f11}$ and $I_{f12} = Y_c V_{f12}$, and $Y_c$ is the line characteristic admittance (i.e., $Y_c = Z_c^{-1}$), (5) can be reformulated only using the sending and receiving voltages of Segment 1 as

$$\begin{bmatrix} V_{f11} \\ Y_c V_{f11} \end{bmatrix} = \begin{bmatrix} A_1 & B_1 \\ C_1 & D_1 \end{bmatrix} \begin{bmatrix} V_{f12} \\ Y_c V_{f12} \end{bmatrix} \quad (6)$$

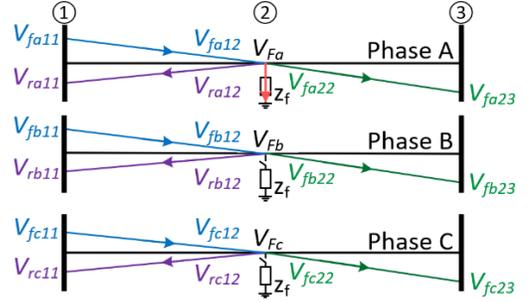

Fig. 3. The propagation of HF waves in faulty TLs.

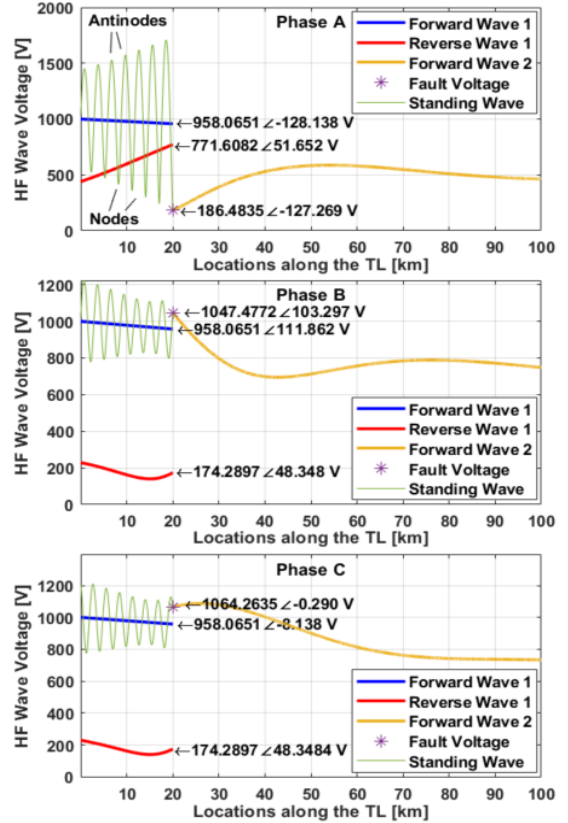

Fig. 4. Typical propagation of HF waves along a TL during SLG faults.

The reverse waves at the ends of Segment 1 are determined by

$$\begin{bmatrix} V_{r11} \\ I_{r11} \end{bmatrix} = \pi_1 \begin{bmatrix} V_{r12} \\ I_{r12} \end{bmatrix} \quad (7)$$

where $V_{r11} = [v_{ra11}, v_{rb11}, v_{rc11}]^T$ and $I_{r11}$ are voltage and current vectors of the reverse waves at the sending end (i.e., Location 1 in Fig. 3), $V_{r12} = [v_{ra12}, v_{rb12}, v_{rc12}]^T$ and $I_{r12}$ are the vectors of reverse voltages and currents at the fault location (i.e., Location 2 in Fig. 3). As $I_{r11} = -Y_c V_{r11}$ and $I_{r12} = -Y_c V_{r12}$, (7) can be written based on voltages as

$$\begin{bmatrix} V_{r11} \\ -Y_c V_{r11} \end{bmatrix} = \pi_1 \begin{bmatrix} V_{r12} \\ -Y_c V_{r12} \end{bmatrix} \quad (8)$$

The voltages at the fault location are the summation of forward and reverse waves. Therefore,

$$V_F = V_{f12} + V_{r12} \quad (9)$$

where, referring to Fig. 3, $V_F = [V_{Fa}, V_{Fb}, V_{Fc}]^T$ is the vector of voltages at the fault location. The relations between the sending and receiving waves at the ends of Segment 2 are

$$\begin{bmatrix} V_F \\ I_{f22} \end{bmatrix} = \pi_2 \begin{bmatrix} V_{f23} \\ I_{f23} \end{bmatrix} \quad (10)$$

where $I_{f22}$ is the vector of forward current waves at the sending of Segment 2 (i.e., Location 2 in Fig. 3), $\pi_2$ is the pi-matrix representing Segment 2, $V_{f23} = [v_{fa23}, v_{fb23}, v_{fc23}]^T$, and $I_{f23}$ are respectively the vectors of receiving voltage and current waves at the end of Segment 2 (Location 3 in Fig. 3). As $I_{f22} = Y_c V_F$ and, $I_{f23} = Y_c V_{f23}$, then

$$\begin{bmatrix} V_F \\ Y_c V_F \end{bmatrix} = \begin{bmatrix} A_2 & B_2 \\ C_2 & D_2 \end{bmatrix} \begin{bmatrix} V_{f23} \\ Y_c V_{f23} \end{bmatrix} \quad (11)$$

When a fault occurs between Segments 1 and 2, the boundary conditions should be determined with respect to the fault type and the engaged phases.

*1) Wave Propagation during SLG, LLG, and 3LG Faults*

Referring to Fig. 3, The boundary conditions at the location of SLG, LLG, and 3LG faults are, as follows:

$$\begin{cases} V_{f12} + V_{r12} = V_F & (12.1) \\ I_{f12} + I_{r12} = I_F + I_{f22} & (12.2) \\ I_F = Y_F V_F & (12.3) \end{cases}$$

where $I_F$ is the vector of fault currents passing through the fault impedances $z_f$ and $V_F = [V_{Fa}, V_{Fb}, V_{Fc}]^T$ is the vector of fault voltage. In the case of an SLG between Phase A and the ground, $I_F$ can be determined by

$$I_F = \begin{bmatrix} V_{Fa}/z_f \\ 0 \\ 0 \end{bmatrix} = \frac{1}{z_f} \begin{bmatrix} 1 & 0 & 0 \\ 0 & 0 & 0 \\ 0 & 0 & 0 \end{bmatrix} V_F \quad (13)$$

and therefore, $Y_F$ is defined as

$$Y_F := \frac{1}{z_f} \begin{bmatrix} 1 & 0 & 0 \\ 0 & 0 & 0 \\ 0 & 0 & 0 \end{bmatrix} \quad (14)$$

In the similar way, $Y_F$ for other fault types can be found, as provided in Table I.

As $I_{f12} = Y_c V_{f12}$, $I_{r12} = -Y_c V_{r12}$, and $I_{f22} = Y_c V_F$,

TABLE I
$Y_F$ FOR DIFFERENT FAULT TYPES

| | Engaged Phases in the Fault | | |
|---|---|---|---|
| | Ag | Bg | Cg |
| SLG | $Y_F := \frac{1}{z_f}\begin{bmatrix}1&0&0\\0&0&0\\0&0&0\end{bmatrix}$ | $Y_F := \frac{1}{z_f}\begin{bmatrix}0&0&0\\0&1&0\\0&0&0\end{bmatrix}$ | $Y_F := \frac{1}{z_f}\begin{bmatrix}0&0&0\\0&0&0\\0&0&1\end{bmatrix}$ |
| | Engaged Phases in the Fault | | |
| | ABg | BCg | ACg |
| LLG | $Y_F := \frac{1}{z_f}\begin{bmatrix}1&0&0\\0&1&0\\0&0&0\end{bmatrix}$ | $Y_F := \frac{1}{z_f}\begin{bmatrix}0&0&0\\0&1&0\\0&0&1\end{bmatrix}$ | $Y_F := \frac{1}{z_f}\begin{bmatrix}1&0&0\\0&0&0\\0&0&1\end{bmatrix}$ |
| | Engaged Phases in the Fault | | |
| | ABCg | | |
| 3LG | $Y_F := \frac{1}{z_f}\begin{bmatrix}1&0&0\\0&1&0\\0&0&1\end{bmatrix}$ | | |
| | Engaged Phases in the Fault | | |
| | AB | BC | CA |
| LL | $Y_F := \frac{1}{z_f}\begin{bmatrix}1&-1&0\\-1&1&0\\0&0&0\end{bmatrix}$ | $Y_F := \frac{1}{z_f}\begin{bmatrix}0&0&0\\0&1&-1\\0&-1&1\end{bmatrix}$ | $Y_F := \frac{1}{z_f}\begin{bmatrix}1&0&-1\\0&0&0\\-1&0&1\end{bmatrix}$ |
| | Engaged Phases in the Fault | | |
| | ABC | | |
| 3L | $Y_F := \frac{1}{z_f}\begin{bmatrix}2&-1&-1\\-1&2&-1\\-1&-1&2\end{bmatrix}$ if fault impedances are delta connected. | | |
| | $Y_F := \frac{1}{3z_f}\begin{bmatrix}2&-1&-1\\-1&2&-1\\-1&-1&2\end{bmatrix}$ if fault impedances are wye connected. | | |

(12.2) can be rewritten as

$$Y_c V_{f12} - Y_c V_{r12} = Y_F V_F + Y_c V_F \quad (15)$$

Solving (12.1) for $V_{r12}$, and applying it into (15) yields

$$Y_c V_{f12} - Y_c (V_F - V_{f12}) = Y_F V_F + Y_c V_F \quad (16)$$

Solving (16) for $V_F$ results in

$$V_F = (Y_F + 2Y_c)^{-1}(2Y_c) V_{f12} \quad (17)$$

Assume that $V_s = [v_{sa}, v_{sb}, v_{sc}]^T$ is the vector of voltage waves generated by the HF wave sources (shown in Fig. 2). As the coupling circuits are equipped with impedance matching units, the amplitudes of the forward voltage waves are one-half of the generated ones at the sending end, $V_{f11} = 1/2 V_s$. By plugging $1/2 V_s$ into $V_{f11}$ in (6) and solving the equation derived from the first row of (6),

$$V_{f12} = 1/2 (A_1 + B_1 Y_c)^{-1} V_s \quad (18)$$

Then, applying (18) into (17) leads to

$$V_F = (Y_F + 2Y_c)^{-1} Y_c (A_1 + B_1 Y_c)^{-1} V_s \quad (19)$$

and finally, by applying (19) into (11), the receiving voltage waves are linked to the wave sources, as follows:

$$V_{f23} = (A_2 + B_2 Y_c)^{-1}(Y_F + 2Y_c)^{-1} Y_c (A_1 + B_1 Y_c)^{-1} V_s \quad (20)$$

As $Y_c = Z_c^{-1}$, (20) can be simplified as

$$V_{f23} = (A_2 + B_2 Y_c)^{-1}(Y_F Z_c + 2I)^{-1}(A_1 + B_1 Y_c)^{-1} V_s \quad (21)$$

*2) Wave Propagation during LL and 3L Faults*

With respect to Fig. 5(a), the boundary conditions for an LL fault between Phases A and B is similar to (12), except for the fault currents that are determined by

$$I_F = \begin{bmatrix} (V_{Fa} - V_{Fb})/z_f \\ -(V_{Fa} - V_{Fb})/z_f \\ 0 \end{bmatrix} = \frac{1}{z_f}\begin{bmatrix} 1 & -1 & 0 \\ -1 & 1 & 0 \\ 0 & 0 & 0 \end{bmatrix} V_F \quad (22)$$

Therefore, $Y_F$ can be derived from (22) as

$$Y_F := \frac{1}{z_f}\begin{bmatrix} 1 & -1 & 0 \\ -1 & 1 & 0 \\ 0 & 0 & 0 \end{bmatrix} \quad (23)$$

when an LL fault engages other phases, $Y_F$ can be found, similarly. $Y_F$'s for different LL faults are given in Table I.

As for 3L faults, the boundary conditions are derived based on Fig. 5(b). The fault current are as follows:

$$I_F = \begin{bmatrix} (V_{Fa} - V_{Fb})/z_f - (V_{Fc} - V_{Fa})/z_f \\ (V_{Fb} - V_{Fc})/z_f - (V_{Fa} - V_{Fb})/z_f \\ (V_{Fc} - V_{Fa})/z_f - (V_{Fb} - V_{Fc})/z_f \end{bmatrix} \quad (24)$$

Reformulating (24) results in

$$I_F = \frac{1}{z_f}\begin{bmatrix} 2 & -1 & -1 \\ -1 & 2 & -1 \\ -1 & -1 & 2 \end{bmatrix} V_F \quad (25)$$

Therefore, it is concluded that in the case of 3L fault,

$$Y_F := \frac{1}{z_f}\begin{bmatrix} 2 & -1 & -1 \\ -1 & 2 & -1 \\ -1 & -1 & 2 \end{bmatrix}. \quad (26)$$

It should be noted that 3L faults are usually shown with wye-connected fault impedances. However, the connection of fault impedances is delta in Fig. 5(b). Therefore, if fault impedances are available in the wye form, those should be converted to the delta form. Hence, $3z_f$ should be utilized in (26), as mentioned in Table I.

## IV. THE FAULT DETECTION ALGORITHM

When a fault occurs, the forward waves are broken into different parts, and only one part continues traveling to the receiving end. Therefore, it is reasonable that some of the receiving waves are decreased during faults. It is mathematically shown in Appendix A that at least one of the receiving waves decreases during faults.

The normalized (i.e., pu) values of the amplitudes of the receiving HF waves are one before the occurrence of faults, as shown in Fig. 6. When faults occur, the grid voltages and currents fluctuate for a while [1], [2], [7], [26], and [27]. Therefore, the HF receiving waves fluctuate for a short time due to the high-frequency variations in the grid, as shown in Fig. 6. Accordingly, the receiving HF waves are invalid during fluctuations. However, once the grid fluctuations with higher frequencies disappear, the receiving waves become valid again, but with varied amplitudes. Based on this sequence, and the fact that at least one of the amplitudes of the receiving waves decrease during faults, Algorithm 1 is proposed for detecting faults, as follows:

In Step 1 of the algorithm, the amplitudes of the receiving waves are measured at any given time $t$.

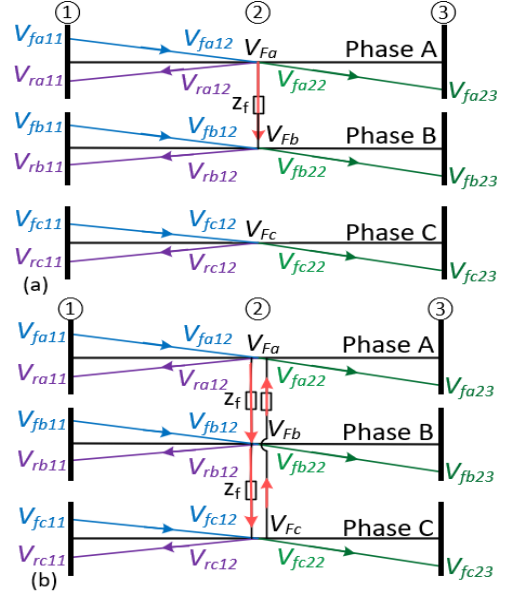

**Fig. 5.** (a) The propagation of the HF waves during LL faults. (b) The propagation of the HF waves during 3L faults.

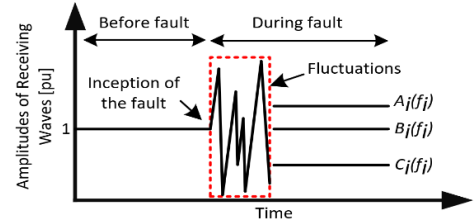

**Fig. 6.** Typical receiving HF waves before and during faults.

---
**Algorithm 1** The Proposed Fault Detection Algorithm.
---
1. Measure the amplitudes of the receiving waves
2. If (amplitudes are valid) → Record time-indexed amplitudes
   Else: Mark the amplitudes as invalid
3. If ($A_i(t), A_i(t-\tau), B_i(t), B_i(t-\tau), C_i(t)$, and $C_i(t-\tau)$) are valid
   {
   If $\begin{cases} A_i(t) < (A_i(t-\tau) - Margin), OR \\ B_i(t) < (B_i(t-\tau) - Margin), OR \\ C_i(t) < (C_i(t-\tau) - Margin) \end{cases}$ → Start Timer
   Else: Stop Timer
   }
4. If Timer ≥ T → Fault
5. Goto 1
---

In Step 2, if the amplitudes are valid, they are time stamped and recorded. Otherwise, the amplitudes are marked as invalid. There are several approaches to assessing the validity of the receiving waves, such as evaluating the quality of the signal or the rate of change of the signal, which are beyond the scope of this manuscript.

In Step 3, it is checked whether the amplitudes of receiving waves at time $t$ (e.g., $A_i(t)$) and their previous values at time $t-\tau$ (e.g., $A_i(t-\tau)$) are valid. If this condition is satisfied, then the amplitudes are compared to their previous values minus a margin. The margin is considered for stabilizing the algorithm against noises or insignificant variations in amplitudes. If any of the receiving waves becomes low, a timer starts counting as long as the receiving wave stays low. The time difference $\tau$ must be larger than $T$, which is explained in the next step, plus a margin for the fluctuations to settle.

In Step 4, if the timer counts for a period of $T$, it indicates the occurrence of a fault. In practice, $T$ should be considered large enough to stabilize the algorithm against noises and other possible variations.

## V. TEST CASES AND DISCUSSION

The grid shown in Fig. 2 is employed as the test system. It can be connected to grids supplied by Synchronous Generators (SGs) or windfarms from either side. The TL is balanced, 345 kV, and 200 km long with the details given in Appendix B. The SG-based grids are modeled with 345-kV voltage sources in series with positive- and negative-sequence impedances of $Z^+ = Z^- = 1.1 + j3.2\ \Omega$ and the zero-sequence impedance of $Z^0 = 5 + j0.5\ \Omega$ based on the Overcurrent Coordination example of EMTP-RV [28]. One 75-MVA windfarm and one 75-MVA load are connected to either side of the TL. The details of the windfarms are available in [29]. The system is operable with any level of IBR contribution.

The frequency of the HF waves sent from Bus 1 (i.e., $f_1$) is 50 kHz. The parameters of line traps and coupling circuits are given in Table II according to [25].

### A. Evaluation of the Proposed Formulation

First, the accuracy of the proposed formulations is evaluated. Different fault types with the impedance of 50 $\Omega$ are applied to the TL at 20 km from Bus 1. The Universal Line Model (ULM), the wideband model in EMTP-RV, used in EMTP-RV for modeling the TL. The trapezoidal rule of integration, with a time step of 1 $\mu$s, is utilized for performing simulations. The simulation results and analytical results based on (21) are given in Table III. The per-unit (pu) values of the receiving waves at Bus 2 are calculated via dividing the during-fault amplitudes by before-fault amplitudes. It is evident in Table III that the analytical and simulation results are almost the same for all faults. The simulation results contain negligible variations which are caused by numerical methods.

In addition, the shaded cells indicate that the amplitudes of the receiving waves on faulty phases are smaller compared to the amplitudes of the waves on unfaulty phases. This is consistent with the descriptions in Section IV.

### B. The Effect of Fault Location

A fault is placed at every kilometer from 1 km to 199 km along the TL. The fault impedance is retained 50 $\Omega$ in all the tests. The pu values of the receiving waves at Bus 2 are shown in Fig. 7. Before the occurrence of faults, the amplitudes of all the receiving waves are 1 pu. However, they differ during faults.

The Ag fault in Fig. 7 shows that the amplitude of the wave on faulty Phase A (i.e., $A_1$) considerably drops for any fault location. It is also observed that $A_1$ becomes drastically small when the fault is close to Bus 2. In contrast to $A_1$, when the Ag fault is close to Bus 2, the amplitudes of the waves on unfaulty Phases B and C (i.e., $B_1$ and $C_1$) grow and even become larger than 1 pu.

As for ABg faults, the amplitudes of both waves on faulty Phases A and B (i.e., $A_1$ and $B_1$) become significantly small during the fault. Similar to Ag faults, $A_1$ and $B_1$ further reduce for faults closer to Bus 2, while the amplitude of the wave on unfaulty Phase C (i.e., $C_1$) increases and pass 1 pu.

TABLE II
COMPONENTS SIZES FOR COUPLING CIRCUIT AND LINE TRAP

| Component | Wave Frequency [kHz] | C [nF] | L [mH] |
|---|---|---|---|
| Coupling Circuit | $f_1 = 50$ | $C_{C1} = 6$ | $L_{C1} = 0.5$ |
| Line Trap | All line traps are narrowband with low and high cutoff frequencies of 45 and 55 kHz | $C_T = 10.31$ [nF] $L_T = 1.5$ [mH] $R_T = 1$ [k$\Omega$] | |

TABLE III
RECEIVING AMPLITUDES AT BUS 2 DURING FAULTS

| SLG Fault | Engaged Phases in the Fault | | | | | |
|---|---|---|---|---|---|---|
| | **Ag** | | **Bg** | | **Cg** | |
| | Rec.† Wave | Amp.‡ [pu] | Rec. Wave | Amp. [pu] | Rec. Wave | Amp. [pu] |
| Analytical Results | $A_1$ | 0.579 | $A_1$ | 0.917 | $A_1$ | 0.911 |
| | $B_1$ | 0.911 | $B_1$ | 0.579 | $B_1$ | 0.917 |
| | $C_1$ | 0.917 | $C_1$ | 0.911 | $C_1$ | 0.579 |
| EMTP Results | $A_1$ | 0.561 | $A_1$ | 0.928 | $A_1$ | 0.898 |
| | $B_1$ | 0.901 | $B_1$ | 0.577 | $B_1$ | 0.908 |
| | $C_1$ | 0.912 | $C_1$ | 0.915 | $C_1$ | 0.562 |
| LLG Fault | **ABg** | | **BCg** | | **ACg** | |
| Analytical Results | $A_1$ | 0.469 | $A_1$ | 0.818 | $A_1$ | 0.469 |
| | $B_1$ | 0.469 | $B_1$ | 0.469 | $B_1$ | 0.818 |
| | $C_1$ | 0.818 | $C_1$ | 0.469 | $C_1$ | 0.469 |
| EMTP Results | $A_1$ | 0.477 | $A_1$ | 0.797 | $A_1$ | 0.459 |
| | $B_1$ | 0.452 | $B_1$ | 0.466 | $B_1$ | 0.788 |
| | $C_1$ | 0.794 | $C_1$ | 0.464 | $C_1$ | 0.465 |
| LL Fault | **AB** | | **BC** | | **AC** | |
| Analytical Results | $A_1$ | 0.518 | $A_1$ | 1.00 | $A_1$ | 0.518 |
| | $B_1$ | 0.518 | $B_1$ | 0.518 | $B_1$ | 1.00 |
| | $C_1$ | 1.00 | $C_1$ | 0.518 | $C_1$ | 0.518 |
| EMTP Results | $A_1$ | 0.517 | $A_1$ | 1.024 | $A_1$ | 0.518 |
| | $B_1$ | 0.521 | $B_1$ | 0.518 | $B_1$ | 1.040 |
| | $C_1$ | 1.061 | $C_1$ | 0.516 | $C_1$ | 0.517 |
| 3L and 3LG Faults | **ABC** | | | | **ABCg** | |
| Analytical Results | $A_1$ | 0.272 | | | $A_1$ | 0.272 |
| | $B_1$ | 0.272 | | | $B_1$ | 0.272 |
| | $C_1$ | 0.272 | | | $C_1$ | 0.272 |
| EMTP Results | $A_1$ | 0.299 | | | $A_1$ | 0.278 |
| | $B_1$ | 0.285 | | | $B_1$ | 0.275 |
| | $C_1$ | 0.278 | | | $C_1$ | 0.274 |

† Rec. stands for receiving. ‡ Amp. Stands for amplitude.

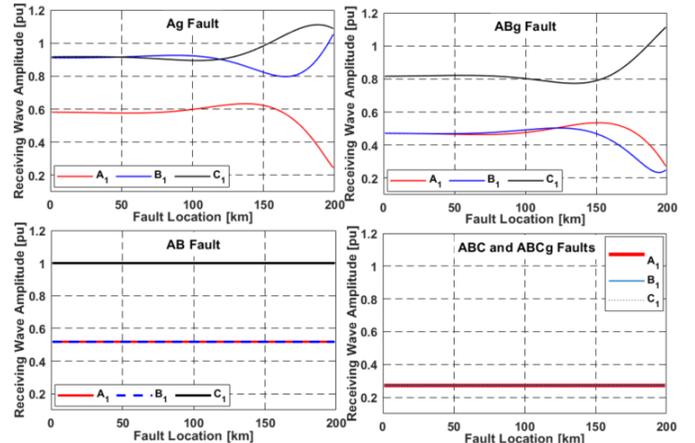

Fig. 7. The amplitudes of receiving waves at Bus 2 for different fault locations and types.

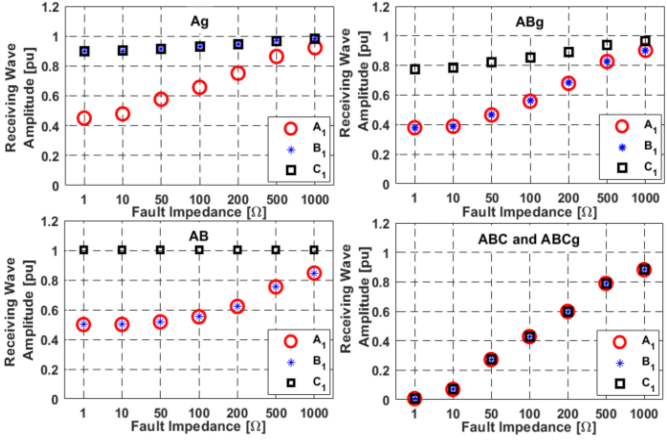

**Fig. 8.** The amplitudes of the receiving waves at Bus 2 for different fault impedances and fault types.

In the case of AB faults, the amplitudes of the waves on faulty Phases A and B equally decrease regardless of the fault location. In contrast, the amplitude of the wave on unfaulty Phase C (i.e., $C_1$) remains 1 pu for any fault location.

The amplitudes of the receiving waves during ABC and ABCg faults are almost the same. In both fault types, the amplitudes equally drop, and their values are not affected by the fault location.

It can be deduced from Fig. 7 that the receiving wave amplitudes are generally smaller for faults engaging more phases. Additionally, at least one of the receiving waves significantly reduces in the case of any fault type at any location along the TL. Therefore, the condition for detecting faults is satisfied.

*C. The Effect of Fault Impedance*

Different fault types are applied to the TL at 50 km from Bus 1. The following fault impedances (i.e., $z_f$'s) are considered for each fault type: 1, 10, 50, 100, 200, 500, and 1000 Ω. The amplitudes of the receiving waves at Bus 2 are shown in Fig. 8.

It is clear that the amplitudes of the waves on faulty phases are reduced more than those on unfaulty phases in all the cases. For example, in the Ag fault, $A_1$ is the smallest amplitude for any fault impedance, or $A_1$ and $B_1$ are smaller compared to $C_1$ in ABg and AB faults.

However, the amplitudes of the receiving waves on faulty phases are larger for higher fault impedances in all fault types. This is predictable since larger fault impedances result in smaller wave reflections. In addition, smaller wave portions flow through higher fault impedances. However, it is clear that at least one of the receiving waves discernably decreases for the fault impedances up to 200 Ω in all fault types. Therefore, the proposed method is able to detect the faults with relatively high impedances.

*D. The Effect of IBR Contribution*

Different faults with the impedance of 20 Ω are applied to 100 km from Bus 1 (the middle of the TL) when the grid is supplied by different IBR contributions, as given in Table IV.

It can be observed in Table IV that there is no obvious difference between the results regardless of the contribution level of IBRs. This is reasonable since the frequencies of electrical parameters and HF waves are significantly different. The receiving HF waves pass through coupling circuits and bandpass filters. Therefore, the electrical parameters at the grid frequency do not affect the receiving waves.

TABLE IV
DURING-FAULT RECEIVING WAVES UNDER DIFFERENT IBR CONTRIBUTIONS

| IBR Contribution | Faulty Phases | | | |
|---|---|---|---|---|
| | Ag | ABg | AB | ABC |
| | Rec. Wave Amp. [pu][†] | Rec. Wave Amp. [pu] | Rec. Wave Amp. [pu] | Rec. Wave Amp. [pu] |
| 100% IBR Contribution | $A_1$ = 0.533 | $A_1$ = 0.421 | $A_1$ = 0.508 | $A_1$ = 0.130 |
| | $B_1$ = 0.924 | $B_1$ = 0.432 | $B_1$ = 0.504 | $B_1$ = 0.137 |
| | $C_1$ = 0.882 | $C_1$ = 0.775 | $C_1$ = 1.03 | $C_1$ = 0.131 |
| 50% IBR Contribution | $A_1$ = 0.541 | $A_1$ = 0.424 | $A_1$ = 0.504 | $A_1$ = 0.135 |
| | $B_1$ = 0.913 | $B_1$ = 0.432 | $B_1$ = 0.505 | $B_1$ = 0.131 |
| | $C_1$ = 0.881 | $C_1$ = 0.779 | $C_1$ = 1.00 | $C_1$ = 0.133 |
| 0% IBR Contribution | $A_1$ = 0.541 | $A_1$ = 0.431 | $A_1$ = 0.503 | $A_1$ = 0.132 |
| | $B_1$ = 0.914 | $B_1$ = 0.432 | $B_1$ = 0.506 | $B_1$ = 0.131 |
| | $C_1$ = 0.882 | $C_1$ = 0.774 | $C_1$ = 1.00 | $C_1$ = 0.132 |

[†] Rec. Wave Amp. [pu] stands for receiving wave amplitude in pu.

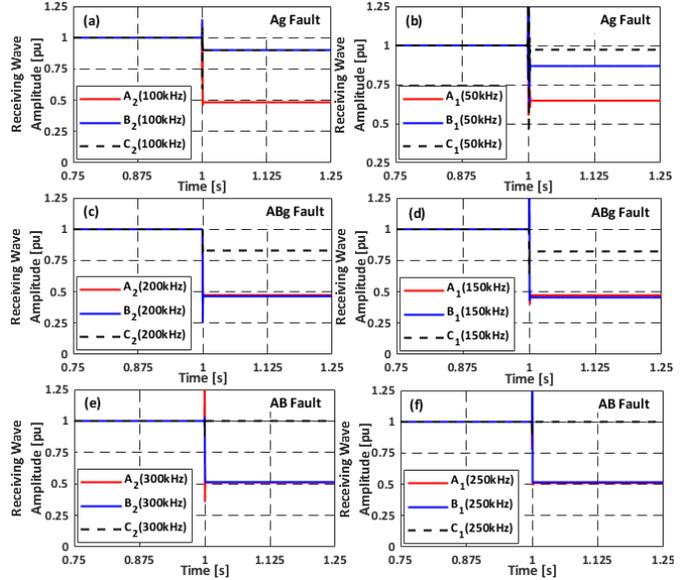

**Fig. 9.** Amplitudes of receiving waves with different frequencies.

In addition, the amplitudes of the waves on the faulty phases are smaller compared to those on the unfaulty phases, as shaded cells in Table IV indicate.

*E. The Effect of Wave Frequency*

In this test, different frequencies are selected for the HF waves sent from Buses 1 and 2. Different fault types with the impedance of 50 Ω are applied to the TL at 50 km from Bus 1 at time $t = 1$ s.

In the case of the Ag fault, the frequencies of the waves sent from Buses 1 and 2 are respectively 50 and 100 kHz. Figs. 9(a) and (b) depict the amplitudes of receiving waves at Buses 1 and 2. It is clear that the amplitudes of the waves on faulty Phase A are drastically reduced at both ends of the line. However, the waves on unfaulty Phases B and C are reduced to a lesser extent, which is consistent with results shown in the Ag cases of Figs. 7 and 8.

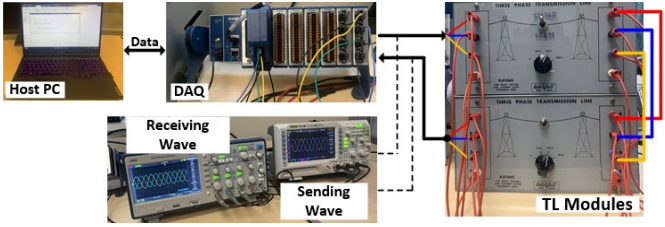

Fig. 10. The setup of the experimental tests.

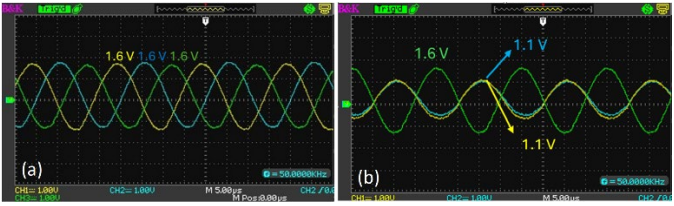

Fig. 11. (a) Observed receiving waves before the inception of the fault. (b) Observed receiving waves during a solid LL fault.

Figs. 9(c) and (d) show the receiving waves for an ABg fault. The frequency of the waves sent from Buses 1 and 2 are 150 and 200 kHz, respectively. The amplitudes of the waves on faulty Phases A and B are reduced more than that on unfaulty Phase C, similar to Figs. 7 and 8. Therefore, the frequencies of the waves do not affect the general behavior of waves during the ABg fault.

In Figs. 9(e) and (f), waves with frequencies of 250 and 300 kHz are respectively sent from Buses 1 and 2. Consistent with the shown waves related to AB faults in Figs. 7 and 8, the amplitudes of the waves on faulty Phases A and B drop at both sides. However, the amplitudes of the waves on solid Phase C remain unchanged at both ends. Accordingly, the frequencies of the HF waves do not affect the general behavior of waves during faults.

*F. Experimental Evaluation*

The experimental setup shown in Fig. 10 is utilized for evaluating the proposed method. Two LabVolt8329 TL modules are sequentially connected, and a characteristic impedance of 120 Ω is selected for both TL modules. One NI-cDAQ9189 bidirectional data acquisition, using NI9262 analog output module, generates HF waves with the frequency of 50 kHz. The amplitudes of waves are 10 V, which is the maximum output voltage of NI9262. The waves are sent through the whole TL, and one NI9223 module is connected to the end of the TL to record the receiving waves. At the same time, two oscilloscopes show the sending and receiving waves to ensure the performance of the test system. The faults are applied at the middle of the line by short-circuiting the terminals of TL modules. It should be noted that the lines are de-energized.

Fig. 11(a) shows the receiving waves before faults. They all have the same amplitude of 1.6 V. However, during a solid LL fault, the amplitudes of the waves on faulty Phases A and B decrease from 1.6 to 1.1 V (i.e., 1 to 0.687 pu) and the amplitude of the receiving wave on unfaulty Phase C remains 1.6 V (i.e., 1 pu), as shown in Fig. 11(b). These results are consistent with the AB faults shown in Figs. 7 to 9.

In addition, different types of solid faults are applied to the middle of the TL, and the receiving waves are recorded and shown in Fig. 12. The variations of the amplitudes in all the cases are similar to the simulation and analytical results shown in Figs. 7 to 9. It should be noted that in the case of ABg and ABCg faults, the poles of the switch are closed with a short time difference. Therefore, a slight time difference is observable in the moments that the amplitudes drop.

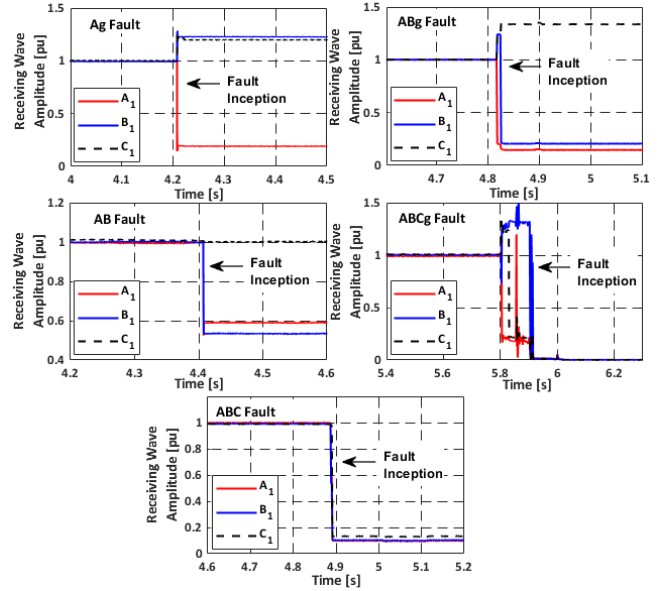

Fig. 12. Experimental measured amplitudes of the receiving waves.

*G. Comparison with Distance Relays*

The performance of the proposed method is compared to distance relays. The generic model of distance relays is utilized in EMTP-RV [30]. The cross-polarization characteristic is selected for the relay, and its Zones 1 and 2 are respectively adjusted to cover 80% and 120% of the TL. The positive-, negative-, and zero-sequence impedances of the TL are $Z^+ = Z^- = 14.914 + 72.3519j$ Ω and $Z^0 = 60.811 + 260.76j$ Ω. The grid is fully supplied by windfarms from both sides. The decoupled sequence control is utilized for the grid side converters of windfarms. Faults with an impedance of 50 Ω are applied to 50 km from Bus 1 at $t = 1$ s.

In the case of the Ag fault, Fig. 13(a) shows that the amplitude of the wave on faulty Phase A drops, and the other two amplitudes reduce to a lesser extent. Therefore, the occurrence of the fault can be detected by the proposed method. However, as Fig. 13(d) shows, the during-fault impedance measured by the relay (i.e., $Z_{Ag}$) is outside Zone 2. Therefore, the relay does not detect the fault.

As for the ABg fault, as Fig. 13(b) depicts, two of the receiving wave amplitudes considerably drop, and the third one reduces less. Thus, this fault is detectable by the proposed method. However, the measured impedance by the relay (i.e., $Z_{ABg}$) is placed inside Zone 2, instead of being in Zone 1, as shown in Fig. 13(d). This leads to the delayed operation of the relay.

In the case of the AB fault, the amplitudes of the HF waves on faulty Phases A and B drop, while the third one remains intact. Therefore, the proposed method can detect the fault. However, the relay-measured impedance (i.e., $Z_{AB}$) is placed close to the border of Zone 2. Therefore, although the relay detects this fault, it reacts to the fault with a delay.

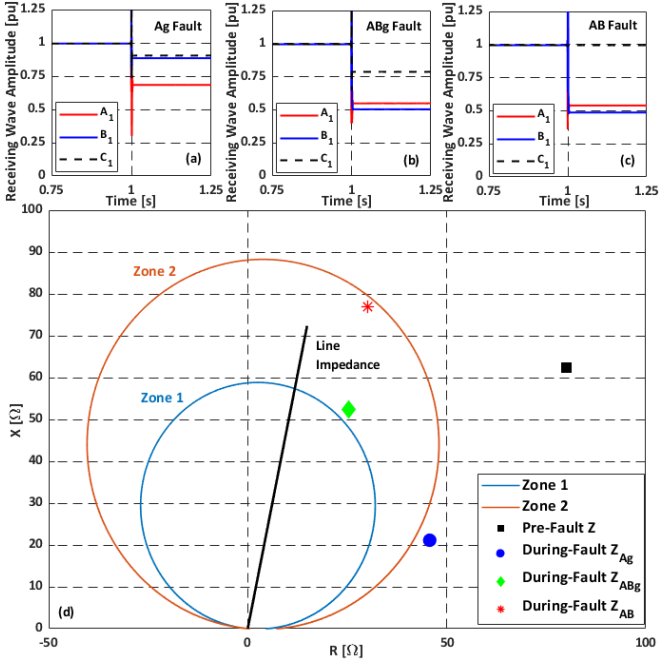

**Fig. 13.** (a)-(c) Measured amplitudes of the receiving waves at Bus 2 for different fault types. (d) Characteristic curve and performance of the distance relay.

## VI. CONCLUSION

The integration of a large number of Inverter-Based Resources (IBRs) into grids can disrupt the performance of protection systems. To address this issue, this paper proposes a fault detection method that utilizes the principles of the propagation of High-Frequency (HF) waves along Transmission Lines (TLs). A set of three balanced HF waves (i.e., the same amplitudes, but 120° phase displacement) are sent from either side of the TL. The frequencies of these two wave sets are different, and they travel along the TL in opposing directions. At the other end of the TL, the receiving waves are measured. It is mathematically shown that faults reduce at least one of the amplitudes. Therefore, a reduction in any of the receiving waves indicates the occurrence of a fault.

The efficacy of the proposed method is evaluated using analytical, simulation, and experimental studies. Simulations are performed using EMTP-RV on an authentic line model. It is concluded from the evaluations that the proposed method is able to detect faults regardless of the type of the sources supplying the grid. The proposed method is also able to detect faults with relatively high impedances.

## APPENDIX A

As $z_f$ is infinitely large before the occurrence of faults, $Y_F = 0$, where $0$ is the zero matrix. By defining $M_1 := (A_2 + B_2 Y_c)^{-1}$ and $M_2 := (A_1 + B_1 Y_c)^{-1} V_s$, (21) can be written as below before faults.

$$V_{f23}^{solid} = M_1 (2I)^{-1} M_2 \quad (A.1)$$

where $V_{f23}^{solid}$ indicates the receiving waves before faults. However, after the inception of a fault, the receiving waves are determined by

$$V_{f23}^{Faulty} = M_1 (Y_F Z_c + 2I)^{-1} M_2 \quad (A.2)$$

where $V_{f23}^{Faulty}$ indicates the receiving waves during a fault.

Based on the triangle inequality [31],

$$\|Y_F Z_c + 2I\| \geq \|Y_F Z_c\| + \|2I\| \quad (A.3)$$

where $\|.\|$ denotes the Frobenius Norm (F-norm), and as norms are positive values and $Y_F Z_c \neq 0$, then

$$\|Y_F Z_c\| + \|2I\| > \|2I\| \quad (A.4)$$

Therefore, it can be deduced from (A.3) and (A.4) that

$$\|Y_F Z_c + 2I\| > \|2I\|. \quad (A.5)$$

According to the norm inequalities [31], the inverse of the terms in (A.5) leads to

$$\|(2I)^{-1}\| \geq \|(Y_F Z_c + 2I)^{-1}\| \quad (A.6)$$

which can be simplified to $\|I^{-1}\| \geq \|(1/2 Y_F Z_c + I)^{-1}\|$. As $\|I^{-1}\| = \sqrt{3}$, and by defining $M := 1/2 Y_F Z_c$, then (A.6) can be written as follows:

$$\sqrt{3} \geq \|(M + I)^{-1}\| \quad (A.7)$$

Based on the Neumann series expansion [31],

$$(M + I)^{-1} = I + \sum_{k=1}^{\infty} (-1)^k (M)^k \quad (A.8)$$

Finding the F-norm of (A.8) and using the triangle inequality,

$$\|(M + I)^{-1}\| \leq \sqrt{3} + \sum_{k=1}^{\infty} \|(-1)^k M^k\| \quad (A.9)$$

If the equality in (A.7) holds, then $\|(M + I)^{-1}\| = \sqrt{3}$, and (A.9) becomes

$$\sqrt{3} = \sqrt{3} + \sum_{k=1}^{\infty} \|(-1)^k M^k\| \quad (A.10)$$

which indicates that $\sum_{k=1}^{\infty} \|(-1)^k M^k\| = 0$. This is impossible since $M \neq 0$, and therefore, $\sum_{k=1}^{\infty} \|(-1)^k M^k\| > 0$. Thus, the equality in (A.7), and consequently, in (A.6) are never held. Therefore,

$$\|(2I)^{-1}\| > \|(Y_F Z_c + 2I)^{-1}\|. \quad (A.11)$$

Considering the Cauchy–Schwartz inequality [31], determining the F-norms of (A.1) and (A.2) results in the below inequalities based on

$$\begin{cases} \|V_{f23}^{solid}\| \leq \|M_1\| \|(2I)^{-1}\| \|M_2\| & (A12.1) \\ \|V_{f23}^{Faulty}\| \leq \|M_1\| \|(Y_F Z_c + 2I)^{-1}\| \|M_2\| & (A12.2) \end{cases}$$

Referring to (A.11), as $\|(2I)^{-1}\| > \|(Y_F Z_c + 2I)^{-1}\|$, then

$$\|V_{f23}^{Faulty}\| < \|V_{f23}^{solid}\| \quad (A.13)$$

This implies that the absolute value of at least one element in $V_{f23}^{Faulty}$ is smaller than the absolute value of at least one element in $V_{f23}^{solid}$. This can mathematically be stated as

$$\exists x \in V_{f23}^{Faulty}, \exists y \in V_{f23}^{solid}: |x| < |y| \quad (A.14)$$

where "∃" means "there exists at least one", "∈" indicates "is an element of", ":" denotes "such that", and |.| is the absolute value. Thus, during faults, at least one of the amplitudes of receiving waves decreases. ∎

## APPENDIX B

The detailed topology of the TL is given in Table BI. The phase conductors are Southwire Cardinal ACSR. The ground conductors are 7-No. 8 based on EMTP-RV Line Database. The soil resistivity is 100 [Ωm].

TABLE BI
THE TOPOLOGY OF THE TL USED IN THE TEST CASES

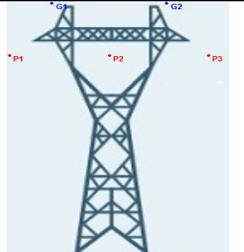

| Conductor | Horizontal Position [m] | Height of Anchor Point [m] | Length of Insulator Chain [m] |
|---|---|---|---|
| P1 | -8.84 | 24.08 | 1.46 |
| P2 | 0 | 24.08 | 1.46 |
| P3 | 8.84 | 24.08 | 1.46 |
| G1 | -5.06 | 28.35 | 0 |
| G2 | 5.06 | 28.35 | 0 |